\documentclass[12pt,preprint]{aastex}

\singlespace

\shorttitle{\map\ Point Sources}
\shortauthors{Wright et al.}

\newcommand{\map}    {{\sl WMAP}}

\newcommand{\lsim}   {\mbox{$_<\atop^{\sim}$}}
\newcommand{\gsim}   {\mbox{$_>\atop^{\sim}$}}

\newcommand{\gt}     {\mbox{$>$}}

\newcommand{\beq}    {\begin{equation}}
\newcommand{\eeq}    {\end{equation}}
\newcommand{\beqa}   {\begin{eqnarray}}
\newcommand{\eeqa}   {\end{eqnarray}}

\slugcomment{submitted to ApJS}

\begin{document}

\title{The Wilkinson Microwave Anisotropy Probe 
(WMAP\altaffilmark{1}) Source Catalog}

\author{
E. L. Wright\altaffilmark{2},
X. Chen\altaffilmark{2},
N. Odegard\altaffilmark{3},
C. L. Bennett\altaffilmark{4},
R. S. Hill\altaffilmark{3},
G. Hinshaw\altaffilmark{5},
N. Jarosik\altaffilmark{6},
E. Komatsu\altaffilmark{7},
M. R. Nolta\altaffilmark{8},
L. Page\altaffilmark{6},
D. N. Spergel\altaffilmark{9,10},
J. L. Weiland\altaffilmark{3},
E. Wollack\altaffilmark{5},
J. Dunkley\altaffilmark{6,9,11},
B. Gold\altaffilmark{4},
M. Halpern\altaffilmark{12},
A. Kogut\altaffilmark{5},
D. Larson\altaffilmark{4},
M. Limon\altaffilmark{13},
S. S. Meyer\altaffilmark{14},
G. S. Tucker\altaffilmark{15}
}

\altaffiltext{1}{\map\ is the result of a partnership between Princeton
University and NASA's Goddard Space Flight Center. Scientific guidance is
provided by the \map\ Science Team.}
\altaffiltext{2}{{UCLA Physics \& Astronomy, PO Box 951547,  Los Angeles, CA 90095--1547}}
\altaffiltext{3}{{Adnet Systems, Inc.,  7515 Mission Dr., Suite A100 Lanham, Maryland 20706}}
\altaffiltext{4}{{Dept. of Physics \& Astronomy,  The Johns Hopkins University, 3400 N. Charles St.,  Baltimore, MD  21218-2686}}
\altaffiltext{5}{{Code 665, NASA/Goddard Space Flight Center,  Greenbelt, MD 20771}}
\altaffiltext{6}{{Dept. of Physics, Jadwin Hall,  Princeton University, Princeton, NJ 08544-0708}}
\altaffiltext{7}{{Univ. of Texas, Austin, Dept. of Astronomy,  2511 Speedway, RLM 15.306, Austin, TX 78712}}
\altaffiltext{8}{{Canadian Institute for Theoretical Astrophysics,  60 St. George St, University of Toronto,  Toronto, ON  Canada M5S 3H8}}
\altaffiltext{9}{{Dept. of Astrophysical Sciences,  Peyton Hall, Princeton University, Princeton, NJ 08544-1001}}
\altaffiltext{10}{{Princeton Center for Theoretical Physics,  Princeton University, Princeton, NJ 08544}}
\altaffiltext{11}{{Astrophysics, University of Oxford,  Keble Road, Oxford, OX1 3RH, UK}}
\altaffiltext{12}{{Dept. of Physics and Astronomy, University of  British Columbia, Vancouver, BC  Canada V6T 1Z1}}
\altaffiltext{13}{{Columbia Astrophysics Laboratory,  550 W. 120th St., Mail Code 5247, New York, NY  10027-6902}}
\altaffiltext{14}{{Depts. of Astrophysics and Physics, KICP and EFI,  University of Chicago, Chicago, IL 60637}}
\altaffiltext{15}{{Dept. of Physics, Brown University, 182 Hope St., Providence, RI 02912-1843}}

\email{wright@astro.ucla.edu}

\begin{abstract}

We present the list of point sources found in the \map\ 5-year maps.  The
technique used in the first-year and three-year analysis now finds 390
point sources, and the five-year source catalog is complete
for regions of the sky away from the galactic plane to a 2 Jy limit,
with SNR $> 4.7$ in all bands in the least covered parts of the sky.
The noise at high frequencies is still mainly radiometer noise, but at
low frequencies the CMB anisotropy is the largest uncertainty.
A separate search of CMB-free V-W maps finds 99 sources of which
all but one can be identified with known radio sources.
The sources seen by \map\ are not strongly polarized.
Many of the \map\  sources show significant variability from year to year, 
with more than a 2:1 range between the minimum and maximum fluxes.

\end{abstract}


\section{INTRODUCTION}
\label{sec:intro}

The Wilkinson Microwave Anisotropy Probe (\map) \citep{bennett/etal:2003}
is a Medium-class Explorer
(MIDEX) mission designed to study cosmology by producing full-sky maps of
the cosmic microwave background (CMB) anisotropy.  
\map\ has measured the angular power spectrum of the CMB anisotropy
over $10^3$ different values of the spherical harmonic index $l$. 
All of these data can be adequately fit by a simple 6 parameter $\Lambda$CDM
model, and this model can also fit other datasets \citep{spergel/etal:2007}.
A determination of the interference from foreground sources is an 
essential part of the analysis of CMB data \citep{nolta/etal:prep}.  
The most important foreground at small angular scales is due to
extragalactic flat-spectrum radio sources.  
Sources are found by searching
the maps for bright spots that approximate the beam profile,
but due to the limited angular resolution of  \map\ it is possible 
to confuse positive CMB excursions with point sources. 
Nonetheless, \map\
provides the only all-sky survey of the millimeter-wave sky so its point source
catalogs are valuable for the study of flat-spectrum radio sources.
In addition, the \map\ point source catalog is used to mask out 
contaminated spots in the high galactic latitude sky used for cosmological
analyses.
208 point sources were found in a search of the first year
of \map\ observations
\citep{bennett/etal:2003c}. 
A search for point sources in the three-year \map\ data found
323 sources \citep{hinshaw/etal:2007}.  In this paper we report on
390 point sources found in the \map\ five-year maps.

The signal to noise ratio on point sources found in WMAP depends 
on the the sensitivity in Janskies per pixel and the number of pixels 
that can be averaged to estimate the flux. Since WMAP was designed 
to give approximately equal sensitivity in each band, and the conversion 
factor from Janskies to Rayleigh-Jeans brightness temperature in 
Kelvins within a constant pixel size is determined by the illuminated 
area of the telescope, the sensitivity in Janskies per pixel is fairly constant.
The $\Gamma_{ff}$ factors tabulated by \cite{hill/etal:prep} give the peak 
temperature expected for a 1 Jansky source 
% with a free-free ($\nu^{-0.14}$) spectrum
as 262.7, 211.9, 219.6, 210.1 \& 179.2 $\mu$K
for the K through W bands of \map.  But the number of pixels that can be averaged
to estimate the flux is proportional to the wavelength
squared, so the overall radiometer noise contribution to
the point source flux uncertainty is 
approximately proportional to the frequency.  \map\
actually illuminates different fractions of the primary
mirror in different bands, and does not have exactly the
same sensitivity in Kelvins per pixel in each band, so the
actual radiometer noise contributions to point source
flux estimates are 0.067, 0.11, 0.13, 0.23 \&  0.40 Jy
divided by the square root of the number of years
of observations for sources on the ecliptic where the coverage is
smallest.  The anisotropy of the CMB itself is also
a source of noise that does not integrate down
with more years of observation.  Using the point-source
flux estimating filters on simulated noise-free CMB maps generated
using the parameters in \citet{spergel/etal:2007} gives $1\sigma$ flux noises
of 0.27, 0.41, 0.36, 0.27 \& 0.14 Jy in the K, Ka, Q, V \& W
bands \citep{chen/wright:2008}.  This ``CMB noise'' term peaks
where the beam size matches the first acoustic peak.

\section{POINT SOURCES IN INDIVIDUAL BAND MAPS}

\label{sec:src}

Extragalactic point sources contaminate the \map\ anisotropy data and a few
hundred of them are strong enough that they should be masked and discarded prior
to undertaking any CMB analysis.  In this section we describe a new direct
search for sources in the five-year \map\ band maps.  Based on this search, we
update the source mask that was used in the five-year analysis. 

In the three-year analysis, we produced a catalog of bright point sources in the
\map\ sky maps, independent of their presence in external surveys.  This process
has been repeated with the five-year maps as follows. We filter the weighted
maps, $N_{\rm obs}^{1/2} T$ ($N_{\rm obs}$ is the number of observations per
pixel) in harmonic space by $b_l/(b^2_lC^{\rm cmb}_l + C^{\rm noise}_l)$,
\citep{tegmark/deoliveira-costa:1998, refregier/spergel/herbig:2000}, where
$b_l$ is the transfer function of the \map\ beam response
\citep{page/etal:2003b, jarosik/etal:2007, hill/etal:prep}, 
$C^{\rm cmb}_l$ is the CMB angular
power spectrum, and $C^{\rm noise}_l$ is the noise power. 
Note that the CMB angular power spectrum used in this filtering has been
updated to match the parameters from the \map\ three-year analysis,
and that the importance of the noise power spectrum goes down as one 
over the number of years of data.
Peaks that are $\gt 5\sigma$ in the filtered map in any band are fit in the 
unfiltered maps for all bands to a Gaussian
profile plus a planar baseline.  The Gaussian amplitude is  converted to a
source flux density using the conversion factors given in 
\citet{hill/etal:prep}.  When a source is identified with $\gt 5\sigma$
confidence in any band, the flux densities for other bands are given if they are
$\gt 2\sigma$ and the fit source width is within a factor of 2 of the true beam
width.  We cross-correlate detected sources with the GB6 \citep{gregory/etal:1996}, 
PMN  \citep{griffith/etal:1994}, and 
\citet{kuehr/etal:1981} catalogs to identify 5 GHz counterparts.  If a 5 GHz
source is within $11\arcmin$ of the \map\ source position (the \map\ source
position uncertainty is $4\arcmin$) we identify the \map\ source with the 5 GHz source
and list the identification in Table \ref{tbl:sources}. When more than one
source lies within the cutoff radius the brightest one is assumed to be the
\map\ counterpart.  

\clearpage
\begin{figure}[tp]
\plotone{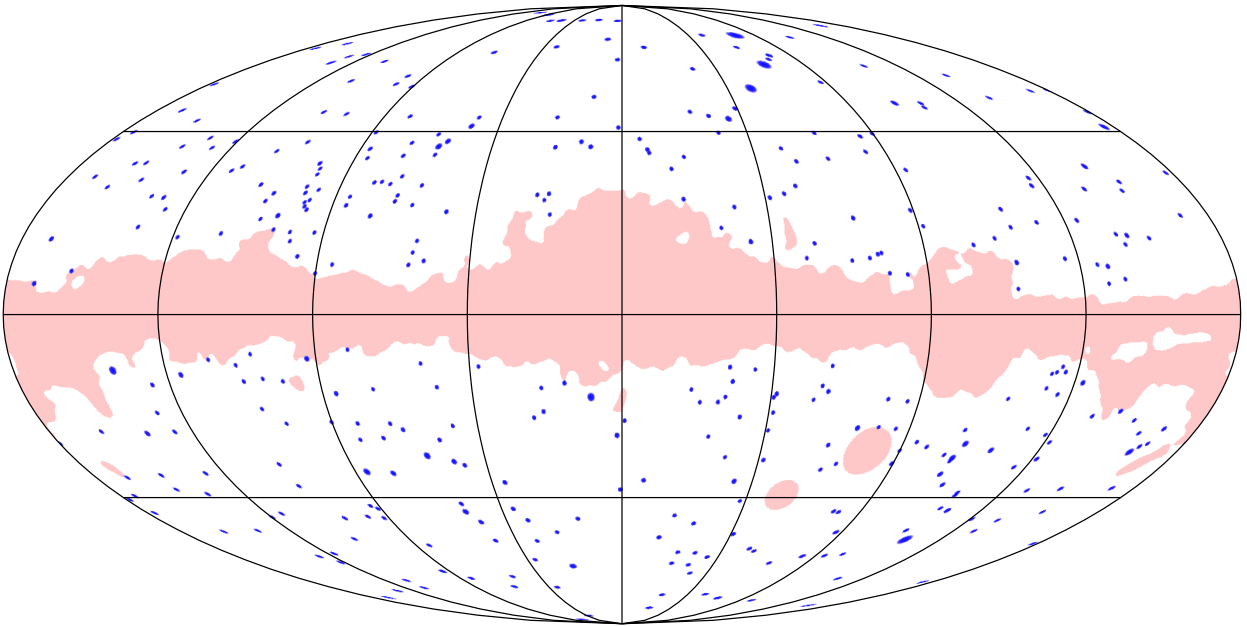}
\caption{Map showing the location of the 390 point sources
found by searching individual band maps.  The shaded region
shows the mask used to exclude
extended foreground emission.  The size of the
plotted points indicates the flux of the source:
the area of the dot scales like the maximum flux over the 5 \map\ bands
plus 4 Jy. Galactic coordinates
are plotted.\label{fig:5yr_ptsrc}}
\end{figure}

\begin{figure}[tpb]
\plotone{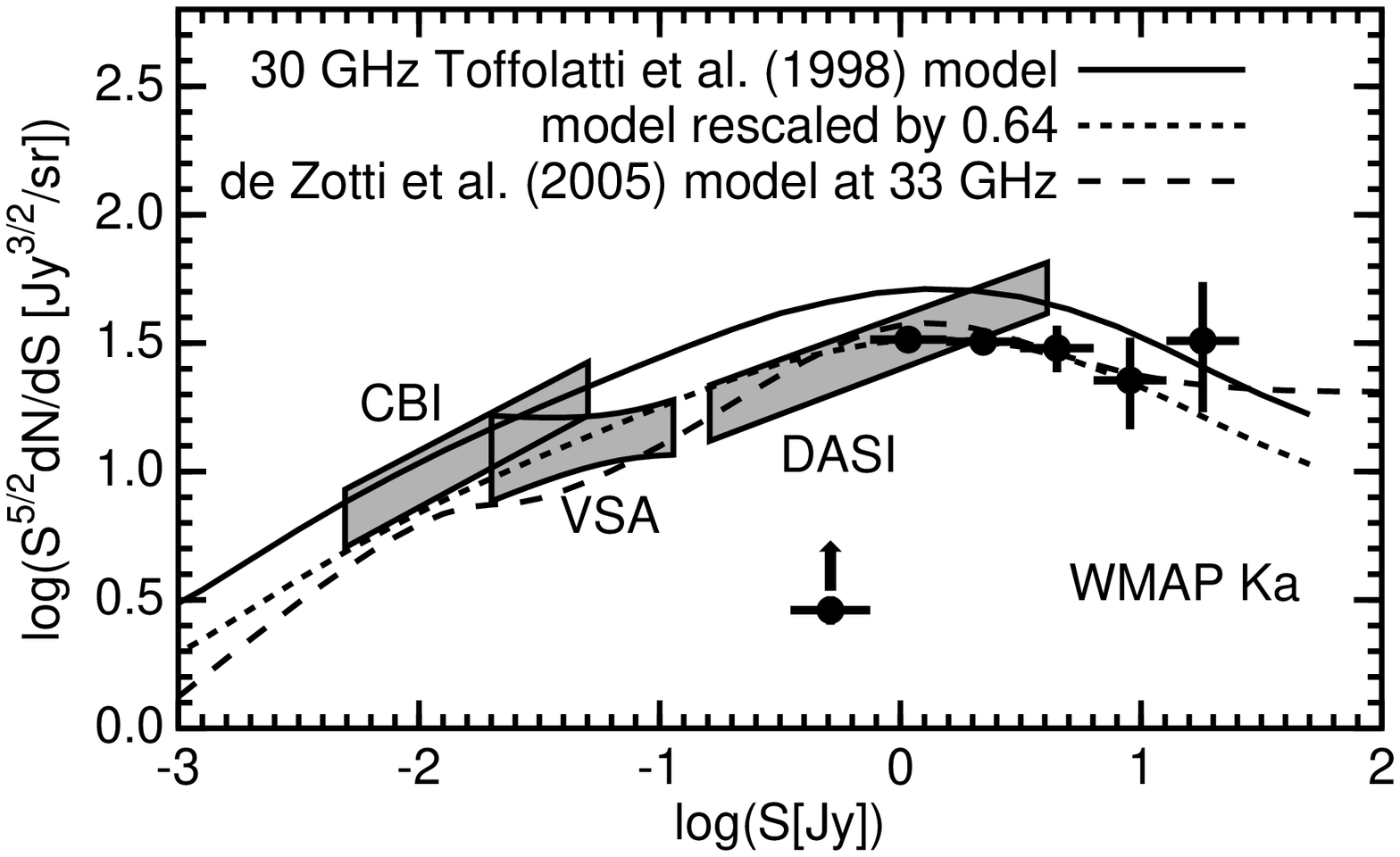}
\caption{Differential source counts from the WMAP five-year
catalog compared the \citet{toffolatti/etal:1998}
model, and to CBI counts at 31 GHz \citep{mason/etal:2003}, 
33 GHz VSA counts \citep{cleary/etal:2005},
and DASI 31 GHz counts \citep{kovac/etal:2002}.  Models
from \citet{toffolatti/etal:1998} and \citet{dezotti/etal:2005} are shown
as well.  
Error bars for \map\ are statistical only.
The \map\ catalog in the 0.35 to 0.75 Jy bin is quite incomplete,
leading to the low data point with the upward arrow on the plot.\label{fig:dNdS}}
\end{figure} 

\begin{figure}[tpb]
\plotone{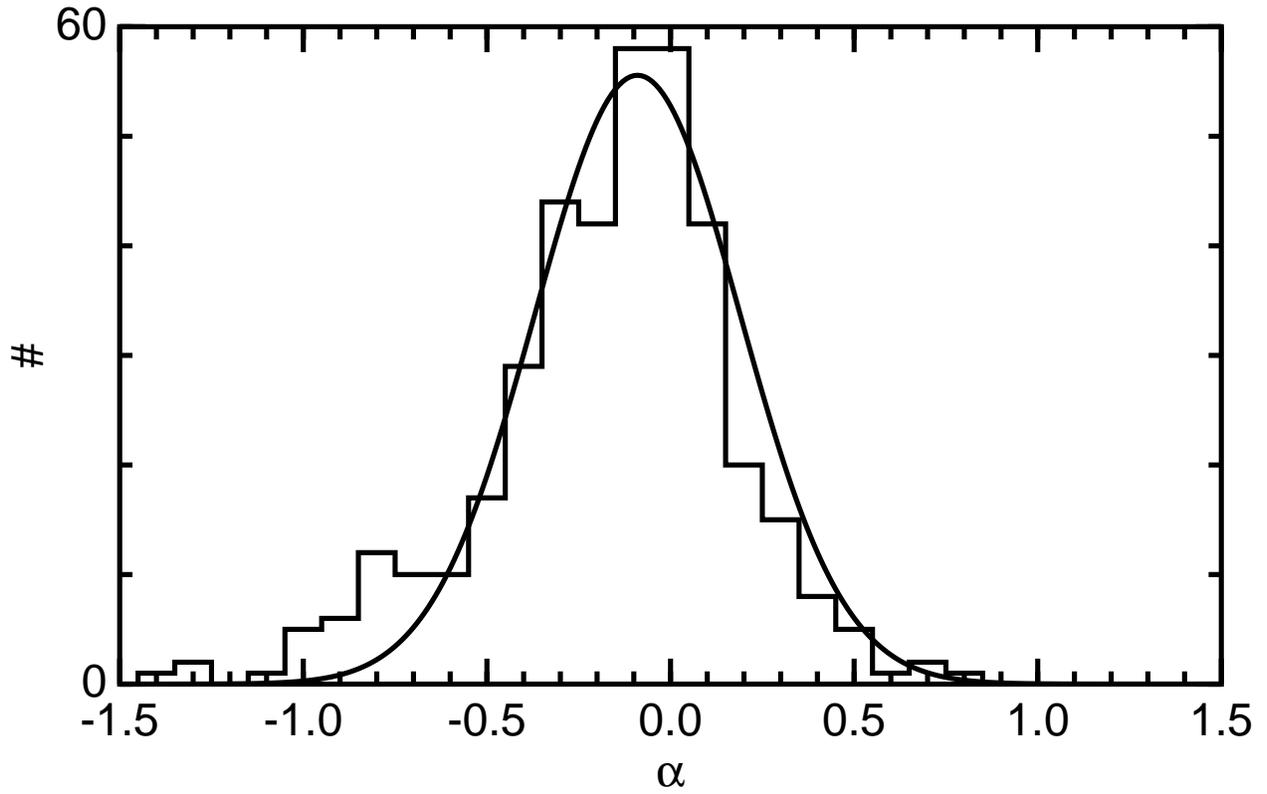}
\caption{A histogram of the spectral indices of \map\  sources
in the five-year maps.  The smooth curve is a Gaussian with
a mean of $-0.09$ and a standard deviation of $0.28$,
normalized to the total number of sources.\label{fig:alpha}}
\end{figure}
\clearpage

The catalog of 390 sources obtained from the five-year maps is listed in
Table~\ref{tbl:sources}.  In the first-year catalog, source ID numbers were
assigned on the basis of position (sorted by galactic longitude).  Now, rather
than assigning new numbers to the newly detected sources, we follow 
\citet{hinshaw/etal:2007} and recommend that WMAP
sources be referred to by their coordinates, e.g., WMAP J0006-0622.  For
reference, we give the first-year source ID in column 3 of
Table~\ref{tbl:sources}.  The 5 GHz IDs are given in the last column. 

The three-year catalog contained 323 sources.  Given the increased sensitivity
in the five-year maps, the number of new sources detected is consistent with
expectations based on differential source count models.  
At the same time, three sources from the first-year catalog are not present in the 
five-year list (numbers 15, 61 \& 156).  Source numbers 31, 96 and 168 which 
were missing in the three-year list have been resurrected.   
Simulations of the first-year catalog
suggested that it contained $5\pm4$ false detections, so the number of
dropped first-year sources is consistent with expectations.
Nine sources from the three-year catalog are missing from the
five-year catalog: WMAP J0513-2015, 0734+5021, 1227+1124, 1231+1351,
1302+4856, 1309+1155, 1440+4958, 1556-7912 \& 1648+4114.  
The sources J1227+1124 and
J1231+1351 were spurious detections caused by sidelobes in the filtered
maps around the strong source J1230+1223.  The problem of strong source
sidelobes is handled as follows in the 5-year analysis:
after identification of each source with 
signal-to-noise ratio greater than 30 in a filtered map, the map is cleaned 
by subtracting the point spread function scaled to the source peak.
A total of 6 out of the 323 sources in the three-year catalog could not be identified 
with 5 GHz counterparts; now 17 out of
the 390 sources in the five-year catalog do not have 5 GHz identifications.
The strong source J1924-2914 is included in the five-year catalog but
not in the previous catalogs because of a small change in the mask used 
to exclude Galactic
plane and Magellanic cloud regions.  Isolated mask regions with fewer than
500 contiguous HEALPix res 9 pixels are no longer included in the mask
(compare Figure \ref{fig:5yr_ptsrc} with the Kp0 mask in Figure 2 of
\citet{bennett/etal:2003c}).  The point source catalog mask shown in 
Figure \ref{fig:5yr_ptsrc} is available on the LAMBDA web site,
http://lambda.gsfc.nasa.gov.

\citet{trushkin:2003} has compiled multifrequency radio spectra and high
resolution radio maps of the sources in the first-year \map\ catalog.  Reliable
identifications are claimed for 205 of the 208 first-year sources.  
Of the 203 sources with
optical identifications, \citet{trushkin:2003} finds 141 quasars, 42 galaxies,
or active galactic nuclei, 19 BL Lac-type objects and one planetary nebula,
IC418.  Forty percent of the sources are identified as having flat and inverted
radio spectra, 13\% might have GHz-peaked spectra, 8\% are classical power-law
sources, and 7\% have a classical low frequency power-law combined with a flat
or inverted spectrum component (like 3C84). \citet{trushkin:2003} suggests that
\map\ source number 116 is likely to be spurious and, for source 61 no radio
component was found.  Indeed, source 61 is not present in either the
three-year catalog or the five-year catalog.
\cite{giommi/etal:2007} observed the 23 objects in the first WMAP 
sample that were not reported as X-ray sources and detected all of these
objects in the 0.3 -10 keV band.  They report a strong correlation between X-ray 
and microwave properties for these blazars.

The distribution of five-year sources on the sky is shown in Figure
\ref{fig:5yr_ptsrc}.  A Kp0+LMC+SMC mask was used when finding
point sources.  This mask excluded 22\% of the sky.
The source counts in the 33 GHz
band are shown in Figure \ref{fig:dNdS}.   The scaling of the
\citet{toffolatti/etal:1998} model has decreased from $0.66$ to $0.64$.
The slope of the WMAP source counts is quite close to the Euclidean
$dN/dS \propto S^{-2.5}$ slope,
while both the models \citep{toffolatti/etal:1998,dezotti/etal:2005}
and the more sensitive data \citep{mason/etal:2003,cleary/etal:2005}
show sub-Euclidean faint source counts.

The spectral indices of the sources are clustered near a flat
spectrum, $\alpha = 0$ in $F_\nu \propto \nu^\alpha$.  A
histogram of the measured $\alpha$'s is shown in
Figure \ref{fig:alpha}.  The smooth curve is a Gaussian
with a mean of $\langle \alpha \rangle = -0.09$ and
$\sigma = 0.28$.  This $\sigma$ includes measurement
errors and is thus an upper limit on the true
dispersion of spectral indices.  Assuming for simplicity that
the underlying distributions of spectral indices is a
Gaussian with standard deviation $\sigma_\circ$,
then the intrinsic dispersion that gives $\chi^2$ per
degree of freedom equal to unity is $\sigma_\circ = 0.176$
and the weighted mean $\langle \alpha \rangle = -0.09$.

\subsection{Analysis of Simulated Maps}

The point source analysis was repeated on simulated maps
constructed with point sources, CMB fluctuations, and
radiometer noise. $10^6$ sources were sampled from
from a power law $N(>S)$ distribution at 30 GHz.  This
distribution was matched to the \citet{dezotti/etal:2005}
source count model.  Spectral indices were then chosen
from a Gaussian with mean -0.09 and standard deviation
0.176, and the fluxes were scaled to the 5 \map\ band
centers.  For each source, the appropriate temperature in each 
band was then added to a randomly
chosen HEALpix pixel at resolution 11
(a total of $12\times4^{11}$ pixels).  These point source
maps, one for each band, were then smoothed with the 
beam window function and converted to a resolution 9 map, 
and added to a simulated CMB plus radiometer noise maps.
The point source detection process was then applied to these 
simulated maps, yielding 363 point sources.  Of these, only 6
were spurious.  The recovered $N(>S)$ agreed with the simulation
input for fluxes $> 1$~Jy, but fell well below the input at lower fluxes.  
Since sources with fluxes $< 1$~Jy are unlikely to be detected,
the ones that are detected tend to have ``benefited'' from a positive
noise or CMB fluctuation, leading to a bias at low fluxes
\citep{eddington:1913}.
The mean ratio of the derived flux to the input flux in bands K-V
is within 5\% of unity for fluxes $> 1$~Jy, but then increases by
10-20\% or more for fluxes $< 1$~Jy.  In the W band the measured
flux is about 10\% below the input flux for fluxes $> 2$~Jy, and
rises to $>20\%$ above the input flux for fluxes $< 1$~Jy.  The bias
in the W band flux for high fluxes could be due to the
Gaussian approximation used in flux fitting.  The deviation of
the mean measured spectral index from the input spectral index is about
-0.02 for Q band fluxes $> 2$~Jy, but rises to $+0.04$~Jy at 1 Jy
and is higher than $+0.10$ for fluxes less than 1 Jy.
We conclude that the fluxes and spectral indices
are reliable for fluxes $> 2$~Jy, but small biases are present for
fluxes $\lsim 1$~Jy.  The source counts should be reliable for
fluxes $\gsim 1$~Jy.

\section{POINT SOURCES IN CMB-FREE ILC MAPS}

The number of sources detected by \map\ as a function of integration
times varied as $N \propto t_{int}^{0.4}$ between the one-year and
the three-year catalogs, but slowed slightly to $\propto t_{int}^{0.37}$ between the
three-year and the five-year maps.  This could be due to the ``noise''
from the CMB, which does not integrate down with increased observing
time.  An approach to circumvent this noise term has been
developed by \citet{chen/wright:2008}.  It involves forming internal
linear combination (ILC) maps from the \map\ bands, but unlike
the normal ILC maps which preserve the CMB and suppress
foregrounds, these ILC maps are designed to suppress the CMB.
Applying this technique to the \map\ V and W bands alone,
\citet{chen/wright:2008} found 31 sources in the one-year maps
% was 63
and 64 sources in the three-year maps.  This gives $N \propto
% was 0.65
t_{int}^{0.66}$ indicating that the ILC technique improves
rapidly with increased observing time.

We have applied this ILC V-W technique to the five-year maps
and there are 99 sources detected in the region with $|b| > 10^\circ$. 
These are listed in Table \ref{tab:ILC-VW}. Among them, 64 are in the WMAP
5 year source catalog,  17 can be identified with sources in NED based on
continuity of spectral energy distributions, 17 are in complex galactic emission
regions, leaving only one source at $09^h21^m28^s$,  
$+7^\circ24^\prime22^{\prime\prime}$ without any identification.  
The V-W technique can find sources sitting in negative peaks of the CMB
where the standard flux finding technique returns an insignificant or even
negative flux.  V band fluxes for these sources have been estimated
by multiplying the value of the V-W map in mK, tabulated in Table  \ref{tab:ILC-VW},
by the median conversion factor derived from the sources 
identified in Table \ref{tbl:sources}.  This factor is 6.28 Jy per mK.
Of the 99 sources in Table \ref{tab:ILC-VW},
% was 13
12 are in the source list by \citet{nie/zhang:2007}
% 8
using the cross-correlation detection method, 8 are in the 
% added line below
are in the new detections of the 
non-blind
catalog by \citet{lopez-caniego/etal:2007}, 27 are in the AT20G Bright Source
% was 73
Sample \citep{massardi/etal:2008}, and 70 are in the CRATES 
catalog \citet{healey/etal:2007}.

The number of sources found by the ILC V-W technique continues
to increase fairly quickly with increased integration time, going like
$t^{0.72}$ from 1 year to 5 years.  For Euclidean source counts
the expected scaling is $t^{0.75}$.

\section{FLUX VARIABILITY OVER FIVE YEARS}

\clearpage
\begin{figure}[tp]
\plotone{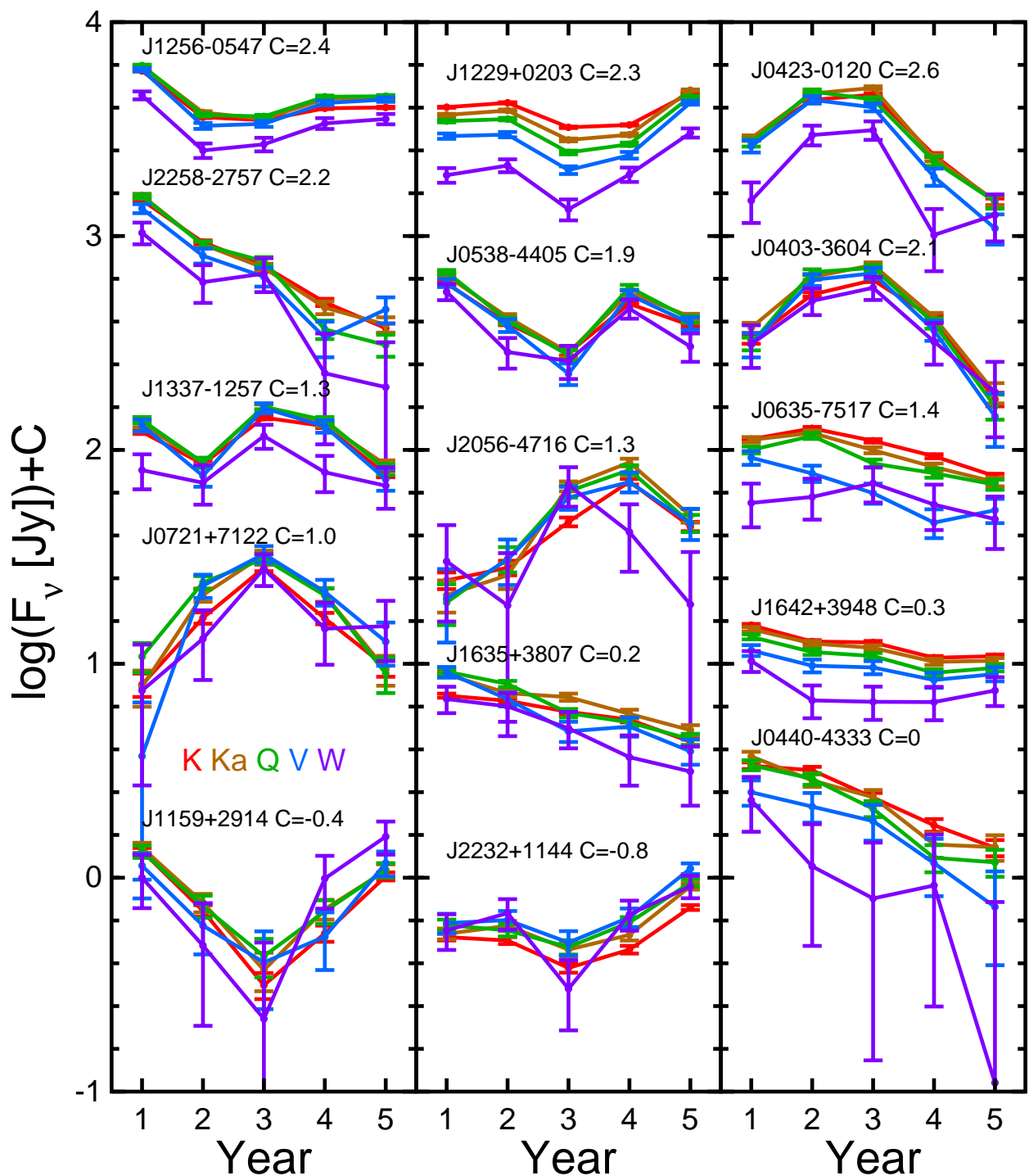}
\caption{The 15 sources with the highest $\chi^2$ for a fit
of a constant flux with an arbitrary spectrum.  The 23 GHz
data are plotted in red, the 33 GHz data are plotted in orange, the 41
GHz data are plotted in green, the 61 GHZ data are plotted in blue,
and the 94 GHz data are plotted in purple.
\label{fig:variables}}
\end{figure}
\clearpage

An analysis of the variability of the \map\ point sources has been performed
by forming fluxes from the individual year maps.  It is possible to measure
the variability of a source without any noise contribution from the CMB
by subtracting the five-year average map from each individual year.  The
fit of a Gaussian beam plus planar baseline to this difference map then gives
a $\Delta F_i$ for the $i^{th}$ year, and the flux for the $i^{th}$ year
is then given by $F_i = \langle F \rangle + \Delta F_i$ where the
five-year average flux is $ \langle F \rangle$.

There are 25 data points for a source detected in all five bands, and
fitting an arbitrary spectrum that is constant in time leaves 20
degrees of freedom.  137 of the 390 sources in Table \ref{tbl:sources}
give $\chi^2  > 37.6$ relative to this fit
and thus are variable at greater than 99\%
confidence.  Sources with $\chi^2 > 37.6$ are flagged with a ``v'' in
the notes column of Table
\ref{tbl:sources}.  The 54 sources with $\chi^2 > 100$ have a ``V''
in the notes column.   These are generally the brighter sources which have
smaller relative flux errors, allowing a better detection of
variability.  The 5 band lightcurves for the 15 sources with $\chi^2 > 450$ 
are plotted in Figure \ref{fig:variables}.
The median rms variability of the Q band fluxes among the 25 brightest Q band
sources is 23\%, after allowing for the flux variations due to radiometer noise. 

It is clear from Figure  \ref{fig:variables} that most of the
variability involves the entire spectrum of a source moving up and
down together, at least on the one year time resolution of this
analysis.  The full table of year-by-year and band-by-band fluxes
for \map\ sources will be available on LAMBDA.

\section{POLARIZATION}

In general the \map\ detected point sources are not strongly polarized.
Of the 390 sources in Table \ref{tab:pol},  only 5 have polarizations
greater than $4\sigma$ in two or more bands.  These sources are listed in
Table \ref{tab:pol}.  In order to assess the average polarization
of the sources, the square of the polarized flux, evaluated as
$Q^2+U^2-\sigma_Q^2-\sigma_U^2$, was fit to the form $p^2 I^2$.
This gave mean polarization percentages of
$p = 2.9$, 2.2, 1.9, $<3.4$ \& $<8.5$\% in K, Ka, Q, V \& W.  
For the V \& W bands 2$\sigma$ upper limits on the mean
polarization percentage are given.

\section{EFFECT ON THE POWER SPECTRUM}

Uncorrelated point sources contribute a power spectrum $C_l =
\mbox{const}$ to the power spectrum.  Since one has to divide by
the beam function $b_l^2$ and multiply by $l(l+1)/2\pi$
to put this on the usual angular power spectrum plot, point sources
give a large contribution to the power spectrum at high $l$.
This can be estimated and removed from the cosmological signal in
several different ways.  The first technique puts an adjustable
constant term in the model $C_l$, while a second technique fits
the difference between frequency bands to a constant $C_l$.  The
CMB gives the same angular power spectrum in different bands, but
the contribution of radio point sources is strongly frequency
dependent: 
\beq C_l^{{\bf i},{\rm src}} = A \, g_i g_{i^\prime}
\left(\frac{\nu_i}{\nu_{\rm Q}}\right)^\beta
    \left(\frac{\nu_{i^\prime}}{\nu_{\rm Q}}\right)^\beta w_l^{\bf
    i},
\label{eq:source_model} \eeq 
where $C_l^{{\bf i},{\rm src}}$ is
the point source contribution to the observed cross-power spectrum
between bands $i$ and $i^\prime$, the factors $g_i$ convert the
result to thermodynamic temperature, $\nu_{\rm Q} \equiv 40.7$ GHz,
and we assume a power law frequency spectrum with index $\beta =
\langle\alpha\rangle-2$.  The window function $w_l^{\bf i} =
b_l^i b_l^{i^\prime} p^2_l$ as in \citet{hinshaw/etal:2007}.
A third technique computes the effect of unresolved point sources
using a model for the counts of sources too faint to be in the
catalog.  This gives \beq C_l = \left(\frac{\partial B_\nu}{\partial
T}\right)^{-2} \int_0^{S_{lim}} S^2 \left(\frac{dN}{dS}\right) dS
\eeq for uncorrelated sources, where $\partial B_\nu/\partial T$
converts temperature into intensity, or equivalently the integral
of $Td\Omega$ in the definition of $a_{l m}$ into flux.  Thus
the point source contribution to an observed cross-power spectrum
can be written 
\beq C_l^{{\bf i},{\rm src}} =
\left(\frac{c^4}{4k^2(\nu_i \nu_{i^\prime})^2}\right) \, g_i
g_{i^\prime} w_l^{\bf i} \int S_i S_{i^\prime} dN \eeq 
where the integral is over all unmasked sources.

If the wrong spectral index is used to convert the difference between power spectra
at different frequencies into a point source contribution, then there will be a
systematic error in the cosmological parameters, primarily in the spectral index
$n_s$.  This effect can be estimated using a simple model for the correction
to the 61 GHz $C_l$ derived from the difference between the 41 and 94 GHz
spectra:
\beq
\Delta C_l^V = \nu_V^\beta \frac{C_l^Q - C_l^W}{\nu_Q^\beta - \nu_W^\beta}
\approx C_l^V(\beta = -2)(1+0.59 (\beta+2) + \ldots )
\eeq
Thus if $\beta$ were really -2.09 instead of -2 then the correction to the 61 GHz
power spectrum should be 5\% smaller than that which would be estimated assuming
$\beta = -2$.  \citet{huffenberger/etal:2006} found that decreasing the point 
source correction by 44\% changed the spectral index $n_s$ by 0.018 so changing 
$\beta$ from -2.0 to -2.09 would change $n_s$ by 0.0022, or $0.15\sigma$.

\section{SUMMARY AND CONCLUSIONS}

There are no other radio surveys that provide the wide coverage of
\map\ at frequencies from 23-100 GHz.  In addition, \map\ provides
year by year fluxes to track the variability of bright millimeter-wave sources.
We present catalogs of point sources found in the \map\ 5 year dataset.
Two different approaches have been used: the standard approach of
looking for peaks in single band maps that have been convolved with
a matched filter, and a new approach that constructs CMB-free
internal linear combination maps.   Using the 61 and 94 GHz data
gives a catalog with somewhat lower sensitivity than the standard approach,
but with better positional accuracy.  The estimated contamination of the
CMB angular power spectrum by unmasked point sources has been estimated,
with results that are consistent with previous analyses and with the
differences between angular power spectra in different bands
\citep{nolta/etal:prep}.  Remaining uncertainties
in the point source correction contribute to the uncertainty of the
cosmological parameters, with the biggest effect occurring for $n_s$.

\acknowledgments
The \map\ mission is made possible by the support of the Science Mission
Directorate Office at NASA Headquarters.  This research was additionally
supported by NASA grants NNG05GE76G, NNX07AL75G S01, LTSA03-000-0090,
ATPNNG04GK55G, and ADP03-0000-092.  EK acknowledges support from an
Alfred P. Sloan Research Fellowship.  This research has made use of
NASA's Astrophysics Data System Bibliographic Services.  We acknowledge
use of the HEALPix, CAMB, and CMBFAST packages.

\clearpage
\setlength{\hoffset}{-20mm}

\begin{deluxetable}{ccccccccccc}
\tablecaption{\label{tbl:sources} WMAP Source Catalog}
\tabletypesize{\footnotesize}
\tablewidth{0pt}
\tablehead{
 \colhead{RA [hms]}
&\colhead{Dec [dm]}
&\colhead{ID}
&\colhead{K [Jy]}
&\colhead{Ka [Jy]}
&\colhead{Q [Jy]}
&\colhead{V [Jy]}
&\colhead{W [Jy]}
&\colhead{$\alpha$}
&\colhead{5 GHz ID}
&\colhead{Notes}
}
\startdata
 00 03 20
&$-$47 52
&   
&\nodata
&$0.7\pm0.06$
&$0.7\pm0.09$
&$0.4\pm0.1$
&\nodata
&$-0.7\pm1$
&\nodata
&
\\
 00 06 06
&$-$06 23
&060
&$2.3\pm0.06$
&$2.3\pm0.1$
&$2.3\pm0.1$
&$2.0\pm0.2$
&\nodata
&$-0.1\pm0.2$
&PMN J0006-0623
&\tablenotemark{v}
\\
 00 10 37
&\phs11 01
&   
&$1.1\pm0.07$
&$1.2\pm0.1$
&$1.2\pm0.1$
&$1.0\pm0.2$
&$1.6\pm0.5$
&$0.1\pm0.3$
&GB6 J0010+1058
&\tablenotemark{V}
\\
 00 12 53
&$-$39 52
&202
&$1.3\pm0.04$
&$1.3\pm0.08$
&$1.0\pm0.09$
&$1.3\pm0.2$
&$0.8\pm0.2$
&$-0.2\pm0.2$
&PMN J0013-3954
&
\\
 00 19 18
&\phs26 03
&   
&$1.0\pm0.06$
&$0.7\pm0.1$
&$0.8\pm0.1$
&$0.5\pm0.2$
&$1.4\pm0.3$
&$0.0\pm0.3$
&GB6 J0019+2602
&
\\
 00 19 40
&\phs20 20
&   
&$1.0\pm0.06$
&$1.1\pm0.08$
&$0.9\pm0.09$
&$1.3\pm0.2$
&\nodata
&$0.1\pm0.3$
&GB6 J0019+2021
&
\\
 00 25 22
&$-$26 02
&   
&$0.9\pm0.05$
&$0.7\pm0.09$
&$0.5\pm0.08$
&\nodata
&\nodata
&$-0.8\pm0.6$
&PMN J0025-2602
&\tablenotemark{a}
\\
 00 26 07
&$-$35 10
&   
&$1.1\pm0.07$
&$1.1\pm0.09$
&$1.4\pm0.1$
&$1.0\pm0.2$
&\nodata
&$0.2\pm0.3$
&PMN J0026-3512
&\tablenotemark{v}
\\
\enddata
\tablenotetext{a}{Indicates the source has multiple possible identifications.}
\tablenotetext{b}{Source J0322-3711 (Fornax A) is extended, and the fluxes listed were obtained by aperture photometry.}
\tablenotetext{c}{Source J0519-0539 is a blend of the Lynds Bright Nebulae LBN 207.65-23.11 and LBN 207.29-22.66.}
\tablenotetext{d}{Source J1356+7647 is outside of the declination range of the GB6 and PMN catalogs.  
Identified as QSO NVSSJ135755+764320 by Trushkin (2006, private communication).}
\tablenotetext{e}{Source J1633+8226 is outside of the declination range of the
GB6 and PMN catalogs. It was identified as NGC 6251 by \citet{trushkin:2003}.}
\tablenotetext{f}{Source J1657+4754 is identified as QSO GB6J1658+4737 by
Trushkin (2006, private communication). Offset from the \map\ position is 18.1
arcminutes.}
\tablenotetext{v}{Probable variable: $\chi^2 > 37.6$.}
\tablenotetext{V}{Variable: $\chi^2 > 100$.}
\end{deluxetable}

\clearpage
\setlength{\hoffset}{0mm}

\begin{deluxetable}{lccccccccc}
\tablecolumns{10}
\tabletypesize{\scriptsize}
\tablewidth{0pt}
\tablecaption{WMAP Point Source Catalog --Five Years VW Bands
\label{tab:ILC-VW}} 						
\tablehead{
\colhead{RA} & \colhead{DEC}  &
\colhead{WMAP ID} & \colhead{Type} & \colhead{Dist.} & \colhead{$f_{V}$\tablenotemark{a}} & \colhead{$T_{V-W}$} & \colhead{5GHz ID} & \colhead{Identified} & \colhead{Note\tablenotemark{c}}\\
\colhead{hms} & \colhead{dms} & \colhead{} &
\colhead{} & \colhead{[arcmin]} & \colhead{[Jy]} & \colhead{[mK]} &  \colhead{} & \colhead{/Masked\tablenotemark{b}} & \colhead{} 
}
\startdata 
\hline
00 06 19 & -06 27 31 & WMAP J0006-0623 & G & 4.1 & 1.7$\pm$0.3  & 0.35 & PMN J0006-0623 & Y / Y & $\ddag$ \\
00 18 49 & 73 24 34 & \nodata  &  QSO & 5.0  & 0.9$\pm$0.3  & 0.23 & GB6 J0019+7327  & Y / Y  & $\ddag$  \\
01 08 43 &  01 39 53 & WMAP J0108+0135 & QSO & 5.0 & 1.8$\pm$0.3  & 0.36 & PMN J0108+0134 & Y / Y & $\ddag$ \\
01 36 55 &  47 50 27 & WMAP J0137+4753 & QSO & 1.2  &  3.1$\pm$0.3  & 0.56  & GB6 J0136+4751 & Y / Y & $\ddag$ \\
02 10 57 & -51 01 28 & WMAP J0210-5100 & QSO & 1.8   &  2.7$\pm$0.3 & 0.42 & PMN J0210-5101 & Y / Y & $\ddag,~\diamond$ \\
02 37 50 &  28 47 49 & WMAP 0237+2848 & QSO & 0.6  &  2.8$\pm$0.3  & 0.51 & GB6 J0237+2848 & Y / Y & $\ddag$ \\
03 19 48 &  41 30 13 & WMAP J0319+4131 & G &  0.5   &  5.6$\pm$0.3 & 1.01 & GB6 J0319+4130 & Y / Y & $\ddag$,~Per A \\
03 21 52 & -37 08 24 & WMAP J0322-3711 & G & 10.6  &  2.1$\pm$0.3   & 0.48  & 1Jy 0320-37 & Y / Y & $\ddag$,~For A \\
\enddata

\tablenotetext{a}{The V-band fluxes of the identified sources are
calculated as in \cite{chen/wright:2008}; The fluxes of the unidentified
sources are estimated by multiplying the V-W temperature in the
filtered map with the median conversion factor from V-W temperatures
to V-band fluxes of the identified sources and are given without
an uncertainty.}
\tablenotetext{b}{Three-year WMAP point source mask is considered here.}
\tablenotetext{c}{$\dag$ and $\ast$ indicate the new sources cross-detected 
in~\citet{nie/zhang:2007} and~\citet{lopez-caniego/etal:2007}.~$\ddag$ 
and $\diamond$ indicate the source is included in the 
CRATES catalog \citep{healey/etal:2007} and the 
AT20G BSS catalog \citep{massardi/etal:2008}, respectively.}
\tablenotetext{d}{Indicates the source has multiple possible 5 GHz
identifications. The brightest one is given here. }
\end{deluxetable}

\clearpage

\begin{deluxetable}{lccc}
\tablecolumns{4}
\tabletypesize{\scriptsize}
\tablewidth{0pt}
\tablecaption{Significant Polarization Percentages
\label{tab:pol}}
\tablehead{
\colhead{WMAP ID} & 
\colhead{K} &
\colhead{Ka} &
\colhead{Q}
}
\startdata
\hline
J0322$-$3711 & $8.5 \pm 0.4$ & $9.3 \pm 1.2 $ & $8.1 \pm 2.2 $ \\
J0519$-$4546 & $5.7 \pm 0.9$ & $8.6 \pm 2.0$ & \nodata \\
J1229$+$0203 & $4.9 \pm 0.3$ & $4.4 \pm 0.6$ & $4.8 \pm 0.8$ \\
J1230$+$1223 & $3.8 \pm 04$ & $4.8 \pm 0.7$ & $4.4 \pm 1.0$ \\
J1256$-$0547 & $3.6 \pm 0.4$ & $3.1 \pm 0.7$ & $3.8 \pm 0.8$ \\
\hline
\enddata
\end{deluxetable}

\end{document}